\documentclass{article}
\usepackage[utf8]{inputenc}
\usepackage[numbers]{natbib}
\usepackage{graphicx}
\usepackage{url}
\usepackage{textcomp}
\usepackage{microtype}
\usepackage{booktabs}
\usepackage{hyperref}
\usepackage{caption}
\usepackage{subcaption}
\usepackage{color}
\usepackage{multirow}
\usepackage[margin=1.2in]{geometry}

\newcommand{\ltb}{\texttt{LazyTensorBarrier()}}
\newcommand{\stf}{Swift for TensorFlow}

\newcommand{\tensor}{\texttt{Tensor}}
\newcommand{\fb}{\\{\small Facebook}}
\newcommand{\brain}{\\{\small Google Research, Brain}}
\newcommand{\cai}{\\{\small Google, Cloud AI}}

\title{LazyTensor: combining eager execution with \\ domain-specific compilers}
\author{
  Alex \c{S}uhan\\{\small Facebook\footnote{Work done at Google Research, Brain}}
  \and
  Davide Libenzi\brain
  \and
  Ailing Zhang\fb
  \and
  Parker Schuh\brain
  \and
  Brennan Saeta\footnote{Correspondence to \texttt{saeta@google.com}}\brain
  \and
  Jie Young Sohn\cai
  \and
  Denys Shabalin\brain
}
\date{\vspace{-2em}}

\begin{document}

\maketitle

\begin{abstract}
Domain-specific optimizing compilers have demonstrated significant performance and portability benefits, but require programs to be represented in their specialized IRs.
Existing frontends to these compilers suffer from the ``language subset problem'' where some host language features are unsupported in the subset of the user's program that interacts with the domain-specific compiler.
By contrast, define-by-run ML frameworks---colloquially called ``eager'' mode---are popular due to their ease of use and expressivity, where the full power of the host programming language can be used.
LazyTensor is a technique to target domain specific compilers without sacrificing define-by-run ergonomics.
Initially developed to support PyTorch on Cloud TPUs, the technique, along with a substantially shared implementation, has been used by \stf\ across CPUs, GPUs, and TPUs, demonstrating the generality of the approach across (1) \tensor\ implementations, (2) hardware accelerators, and (3) programming languages.
\end{abstract}

\section{Introduction}
\label{section:introduction}

Imperative, sequential program execution---colloquially called ``eager execution'' or ``define-by-run''~\cite{chainer} in ML contexts---is easily understood and expressive, which is why it is used as the basis for most widely-adopted programming languages.
Popular libraries for machine learning centered around eager execution such as PyTorch~\cite{pytorch} and NumPy~\cite{numpy} are known to be both flexible and easy to debug.
Programs using these libraries dispatch ``kernels''---pre-compiled functions such as matrix multiplication, convolution, or element-wise arithmetic operations on \texttt{Tensor}s ($n$-dimensional arrays)---to computational devices (e.g. CPUs or GPUs).

On the other hand, optimizing domain-specific compilers (DSCs) \cite{xla, tvm, glow} substantially improve performance of machine learning models.
Additionally, these compilers are sometimes the \emph{only} way to target domain-specific accelerators, such as Cloud TPUs~\cite{cloudtpu}.
The downside: a user's program must be presented to these DSCs in a compiler-specific intermediate representation (IR).
Because these IRs are focused on a particular domain, they typically do not aim to be as expressive as general-purpose programming languages.
While numerous libraries in general-purpose programming languages have been developed to build these IRs, they all suffer from the \emph{language subset problem} where expressivity is sacrificed in the portion of the user's program that uses the library to align with the capabilities of the target IR.

In this paper we introduce \emph{LazyTensor}\footnote{Unrelated to \texttt{LazyTensor} in \url{https://gpytorch.ai/}}, a novel approach to combine the benefits of eager execution with DSCs.
Our technique allows full use of all host programming language features throughout the \tensor\ portion of the user's program, avoiding the language subset problem.
Initially developed to support PyTorch on Cloud TPUs, the LazyTensor approach has been adopted by two numerical computing libraries in two different programming languages while sharing the majority of the implementation.
The main contributions of this paper are:

\begin{enumerate}
    \item 
    A technique for combining an eager programming model of \tensor\ programs with domain specific compilers that does not restrict the expressivity of the user's programming language.
    The approach is general enough to be applied to any define-by-run machine learning framework.
     (Section~\ref{sec:idea})
    \item
    An implementation of the LazyTensor design across two different machine learning frameworks in two different programming languages: PyTorch and Swift for TensorFlow. (Section~\ref{sec:impl}) 
    \item 
    An evaluation of our design across multiple languages, \tensor\ types, and accelerators (GPUs and TPUs). (Section~\ref{section:evaluation})
\end{enumerate}

\section{Background}
\label{sec:ctx}

Deep learning models are often trained using libraries centered around a multi-dimensional array abstraction, often called a \tensor~\cite{numpy, pytorch, tensorflow}.
The model is (1) a collection of \texttt{Tensor}s corresponding to learned parameters (weights) and (2) a sequence of operations mapping (a) the parameter collection and (b) input \tensor(s) (e.g. a batch of images represented as a 4-dimensional array\footnote{The \tensor\ can be arranged as [batch $\times$ image height $\times$ image width $\times$ image channels].}) to the output \tensor\ result (e.g. a 2-dimensional\footnote{The shape corresponds to: [batch $\times$ \#-classes].} \tensor\ corresponding to a one-hot encoding classification category per image).

Domain-specific optimizing compilers have been developed around the \tensor\ abstraction to (a) target domain-specific hardware such as TPUs, and/or (b) eke the maximum performance out of a given hardware footprint.
These domain specific compilers consume source programs as input in compiler-specific intermediate representations (IRs), e.g. XLA HLO IR~\cite{xla}.
These IRs are not designed to be human authored, as they either have extremely verbose syntax, or inflexibility (e.g. no dynamic memory allocation, and require static-shapes for every \tensor).

\subsection{Graph Mode}

Historically, many deep learning frameworks \cite{tensorflow, caffe, theano} represent models as data structures.
The TensorFlow v1~\cite{tensorflow} system is organized around the construction and execution of dataflow graphs.
A TensorFlow Python program acts a meta-program, where the Python code builds a computational graph.
This graph is handed off to the dataflow execution engine, implemented in C++.
Because this graph can be serialized as a GraphDef protocol buffer, and because it executes independently of the host programming language, we refer to it as the GraphDef programming language, and the dataflow execution engine as the interpreter of the GraphDef programming language.
Because the GraphDef programming language is optimized for \tensor's and does not support many of Python's features,\footnote{Including exceptions, reflection, dynamic dispatch, and more.} it can be translated into an equivalent program in a DSC's IR.

\subsection{Eager programming \& asynchronous execution}

In contrast to graph mode, define-by-run libraries \cite{chainer, pytorch}---where the neural network model is written in code and directly executed---are seen as easier to debug and more flexible, because users have the full power and expressivity of a general purpose programming language.
Compute hardware (CPUs as well as accelerators like GPUs) are efficiently utilized through asynchronous execution.
When a user's program requests a matrix multiplication between two \tensor s, the operation's ``kernel'' is dispatched to the compute device, and control is immediately returned to the user's program.
The runtime blocks the user's program only when it attempts to view the contents of a \tensor.
One can think of the \tensor\ type as a \textit{promise} to return a concrete \tensor\ value  at some future time.
Because the \tensor\ computation is never materialized in a data structure, there is not a program representation that can be translated to a DSC's IR.

\subsection{Eager programming \& DSCs} %

There are a number of mechanisms for staging code out of an eager \tensor\ programming environment so that it can be optimized by a DSC.

\textbf{Tracing}.
One set of methods involves a user-directed mechanism for running eager program code, with specialized \tensor\ types or in a specialized context, while recording information about what \tensor\ operations were executed.
This tracing process is akin to that used by operator-overloading reverse-mode automatic differentiation systems like Autograd~\cite{autograd-thesis} to build a ``tape'' that is walked backwards during gradient evaluation.
Systems like JAX\footnote{An acronym for, among other things, ``JAX is Autograd and XLA''.}~\cite{jaxtracing} and TensorFlow~2~\cite{tensorflow-eager} provide tracing decorators that cause annotated functions to be executed in a context that abstracts \tensor s into \emph{representations} of potential values and captures \tensor\ operations for optimization by a DSC.
On its own, tracing with abstract values is essentially a more-ergonomic interface for a ``graph mode'', while host language features like control flow and all non-\tensor\ code are ``invisible'' to tracing and either are executed only at trace time or, if dependent on runtime \tensor\ values, cannot be traced at all.

\textbf{Language virtualization}.
Tracing can be augmented by a source-to-source transformation that rewrites untraceable language features like control flow into traceable ones.
This ``virtualization'' process is exemplified by the Lightweight Modular Staging (LMS)~\cite{lms} framework for Scala and by AutoGraph~\cite{autograph}, a Python virtualization system adopted by TensorFlow 2.
Once augmented with virtualization, tracing systems are able to stage out eager code even if it uses operations on built-in types like Python lists whose behavior can't be overloaded, or language-native control flow that branches on runtime \tensor\ values.
Virtualization may not cover all language features, with exception-based control flow forming an especially difficult case.

\textbf{Direct compilation}.
Another approach to bridging eager code and a DSC is to implement a compiler frontend. TorchScript implements a lexer, parser, type inference engine, and optimizer for a \tensor--focused subset of Python that includes most real-world PyTorch code, while Julia's XLA support~\cite{juliaTPU} leverages the Julia compiler to do the same for a subset of that language.
When embedded in dynamic languages like Python and Julia, such a compiler can be invoked just-in-time, with a user interface similar to tracing (e.g. a function decorator).
Like tracing, approaches based on a compiler frontend typically require that all code in the staged-out function either be statically evaluated\footnote{If this static evaluation is not performed, or performed using an alternative language implementation (as in TorchScript), some language features may not be supported at all.} at compile time or be present in the final program handed off to the DSC.
This restriction is viral: compilation of one user function requires compilation of functions that it calls, so unsupported behavior in a function or library (e.g., a call to a foreign function like a physics simulator or environment model) means that all \emph{transitive} callers also cannot be compiled.

A key common downside to these existing techniques is that they only support a subset of the host programming language.
Additionally, these approaches prevent the user from interleaving code that should be compiled with a DSC and code that shouldn't.
This is similar to the ``function coloring problem'' from async-await contexts \cite{functionColor}; functions not compiled with a DSC can call functions compiled with a DSC, but not the other way around.

\section{The LazyTensor Approach} %
\label{sec:idea}

Types in common general-purpose programming languages operate with eager semantics; because our approach maintains the \emph{illusion of eager execution}, our approach does not compromise the host programming language.
In particular we support the complete host language, including interleaving arbitrary \tensor\ and non-\tensor\ computations.

Our approach build upon the insight behind PyTorch's asynchronous execution: as long as the user program does not observe a \tensor's contents, the user's program cannot distinguish when a \tensor\ operation is actually executed.\footnote{Ignoring timing and similar side-channels.}
Instead of dispatching each op individually to execute asynchronously, we buffer sequences of operations in a data structure.
Additionally, we observe that these operation sequences can be transformed into a DSC's IR.
The programs generated by a DSC are semantically equivalent to running the operations individually, and thus we can replace executing the operation sequence with direct invocation of the compiled DSC program.
\emph{\texttt{Tensor}s are neither promises nor representations of future data, but both simultaneously!}

We call these ``sequences of operations'' an \emph{IR graph}.
The IR graph construction process always begins from a \tensor\ operation.
This reflects the separation between the host language and the LazyTensor library: calls to \tensor\ operations are the only entry points to our system.
The first such \tensor\ operation is always a ``factory method'', which creates a \tensor\ from an existing \tensor, a range, a repeated constant value, or random values sampled from a specified distribution.

The entire \tensor\ API can be divided into two domains, operations that can be represented in a IR graph (\emph{IR compatible} operations), and operations that force the evaluation of a IR graph (\emph{IR incompatible} operations).
Any \tensor\ operation that exclusively returns one or more \tensor's fall into the first domain.
Examples include matrix multiplication, convolution, and element-wise arithmetic.
Operations on \tensor s that return non-\tensor\ types are incompatible with IR construction, such as operations that return a scalar value (e.g computing a string-representation to print to a console, or a boolean to make data-dependent control-flow decisions in the host programming language).

The entire \tensor\ API is available at all points in the user's program. 
This has a number of advantages including the ability to use the complete host language for:

\begin{enumerate}
    \item \textbf{Function abstraction.}
    IR graph recording happens transparently through host-language abstractions such as functions.
    We need not differentiate functions which are incompatible with the IR graph, and thus avoid the function coloring problem.
    Any function can call any other function, irrespective of whether it is composed exclusively of IR compatible operations.
    Functions that call IR-incompatible operations simply force the evaluation of the IR graph at runtime, and a new IR graph is subsequently started.
    \emph{Changes to the implementation of one function never affect the ability to optimize any other function.}
    
    \item \textbf{Control flow.}
    Because we maintain an eager API, the complete set of host language control flow mechanisms function identically to an eager implementation.
    This includes all host language control-flow operations including complicated cases such as exceptions or virtual function calls.
    
    \item \textbf{Data structures.}
    Programs can embed \tensor\ values as part of arbitrary data structures. 
    This allows for composition with non-\tensor-aware libraries which could embed \tensor\ values as boxed (in dynamic languages) or type-specialized first-class values (in compiled languages).
\end{enumerate}

As a result, in contrast to prior staging systems, we make no compromises on the integration with the host language.
We achieve this by effectively building a sophisticated version of the eager runtime that relies on recording an IR graph behind-the-scenes, rather than exposing it to the user.

\section{The LazyTensor System} %
\label{sec:impl}

The LazyTensor system builds on top of (a) an underlying eager runtime (either PyTorch or TensorFlow-Core) and (b) the XLA domain-specific compiler.
LazyTensor has 3 main components: (1) a custom \tensor\ type with an identical API to an existing \tensor\ type, (2) a mapping from the high-level \tensor\ operations to XLA HLO sequences implementing the semantics of the requested operation (called a ``lowering''), and (3) a runtime that lowers sequences of \tensor\ operations into XLA HLO IR, and orchestrates compilation \& execution of the resulting program.
Because compilation often is very expensive, the LazyTensor system carefully caches and re-uses program IR graphs keyed on a canonicalized form.

The LazyTensor implementation includes an additional ``barrier'' API (\texttt{mark\_step()} in PyTorch, \ltb\ in \stf).
This API completes the current in-progress IR graph construction, and dispatches it to the runtime for compilation and execution.
The barrier API takes a boolean parameter to control whether the call should block until IR graph execution has completed and all \tensor\ data has been materialized in memory.
Implementations of IR incompatible operations call the barrier API with the blocking bit set before proceeding with their implementation.
Future work can remove the barrier API from the public interface.

The barrier needs information regarding all live \tensor s for which the outstanding computation accumulated for the step needs to be executed to set their value.
The liveness information is tracked per device by a context object, \texttt{DeviceContextArena}, available for the entire lifetime of the process.
Constructors and destructors on the \tensor\ type call \texttt{RegisterTensor} and \texttt{UnregisterTensor} methods (respectively), using unique identifiers generated when tensors are created.

The core of the LazyTensor system is implemented in C++, and has been substantially shared across both PyTorch and \stf~\cite{s4tf}.
The custom \tensor\ types are in Python and Swift (respectively), and are thus not shared.
Further, because the \tensor\ APIs are not identical, the mappings from user-level operations to XLA HLO are tweaked accordingly.

The LazyTensor system has a number of other features to make the system useful in practice.
LazyTensor supports distributed training, including leveraging the custom high-speed chip-to-chip network on TPU Pods by exposing the relevant XLA HLO collective operations (e.g. cross-replica-sum).
Finally, LazyTensor adds support for automatic mixed precision~\cite{automaticMixedPrecision} enabled by an environment variable.

\subsection{IR graph representation}

\begin{figure*}
\centering

\begin{subfigure}[b]{0.45\textwidth}
\begin{verbatim}
IR {
    %0 = f32[] prim::Constant(),
        value=1
    %1 = f32[2,4] aten::expand(%0),
        size=(2, 4)
    %2 = f32[2,4] xla::device_data(),
        device=CPU:0
    %3 = f32[2,4] aten::mul(%2, %1)
    %4 = f32[2,4] xla::device_data(),
        device=CPU:0
    %5 = f32[2,4] xla::device_data(),
        device=CPU:0
    %6 = f32[2,4] aten::mul(%5, %4)
    %7 = f32[2,4] aten::add(%6, %3),
        ROOT=0
}
\end{verbatim}
\caption{Textual form}
\end{subfigure}
\begin{subfigure}[b]{0.45\textwidth}
    \centering
    \def\svgwidth{0.9\columnwidth}
    \begin{tiny}
    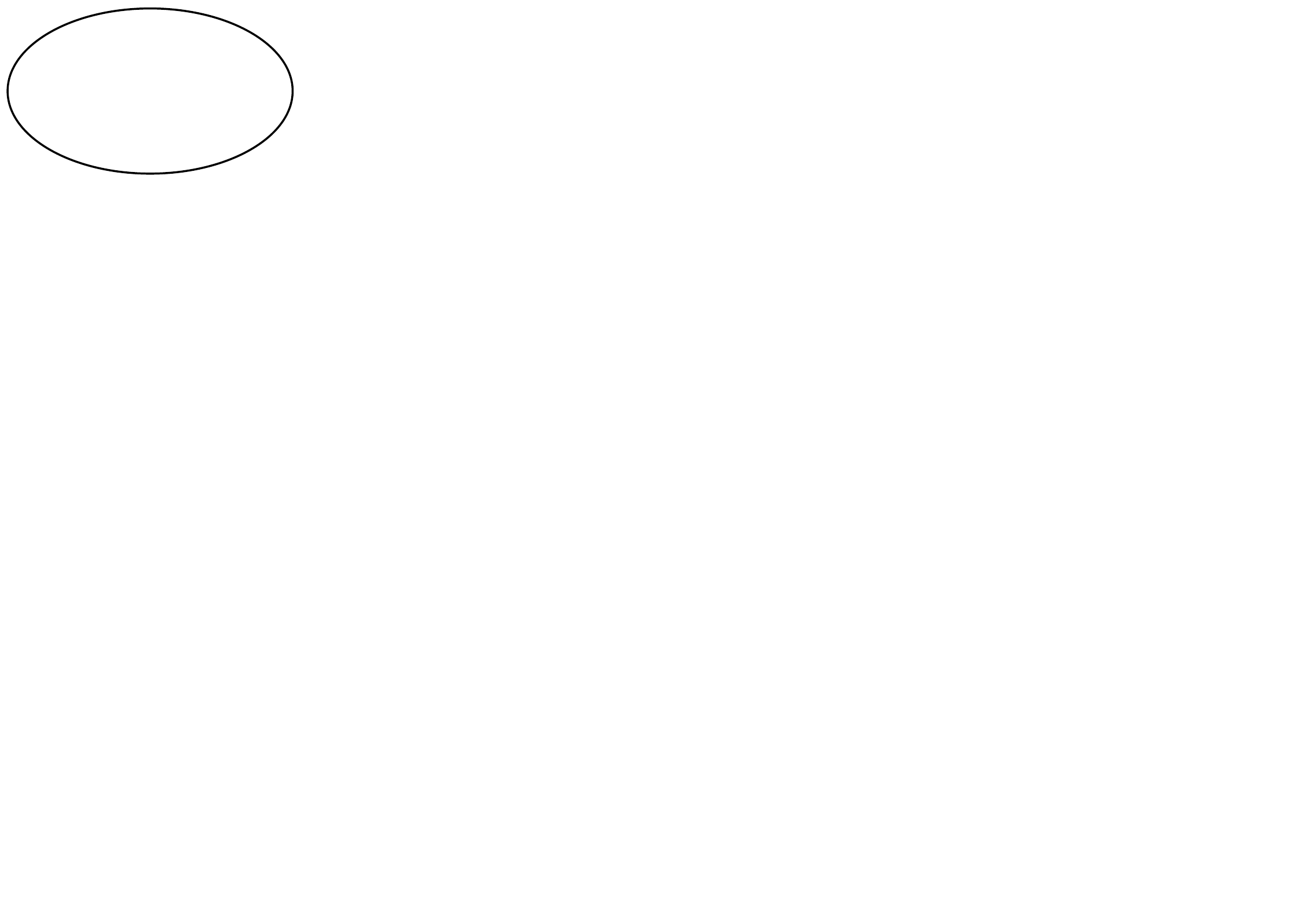
    \end{tiny}
    \vspace{1.5cm}
    \caption{Visual form}
\end{subfigure}
\caption{An IR graph of \texttt{x * y + z}.}
\label{fig:pysmall}
\end{figure*}

The LazyTensor IR graph records a computation as a directed acyclic graph, in which the leaves are the inputs and the roots are the results computed based on the given inputs.
Figure~\ref{fig:pysmall} contains two representations of the IR graph constructed from PyTorch for a simple \texttt{x * y + z} computation, on floating point tensors of shape \texttt{[2, 4]}.

All the nodes record their shape in addition to the operation type itself, in this case \texttt{f32[2,4]} for the shapes, with the exception of the scalar constant 1 which is represented as rank zero: \texttt{f32[]}.
Inputs are represented as \texttt{xla::device\_data}.

The multiplication and addition are represented by nodes \texttt{\%6} and \texttt{\%7}.
In PyTorch, the addition operation allows specifying a scaling factor for the second parameter, which is represented by the additional node \texttt{\%3}.
The scaling factor is the constant 1 expanded to the required shape.
The scaling operation is optimized away by the underlying XLA compiler backend.
The generated native code will not materialize the constant nor execute the multiplication by 1.

Optimizing away operations which involve known scalar values must be balanced against the ability to reuse previously-compiled code for IR graphs which differ only in such values.
To achieve the latter, scalar values could instead be treated as computation parameters.
For example, certain classes of models involves indexing at a variable starting position and a fixed length---compiling a separate version for each starting index doesn't improve performance on any class of hardware and the user experience is degraded by the increased compilation times.
We've chosen a simple heuristic: treat 0 and 1 as special scalars, which have their values encoded in the IR graph while treating the rest as dynamic computation parameters.
This covers the elision of multiplication by 1 or adding zero-initialized gradients that result from lowering some high level APIs.
While this might trigger spurious recompilations due to incidental occurrences of special constants, we haven't observed this to be a problem in practice.

Analogously to native PyTorch, there are no implicit transfers between devices.
This is reflected in the IR graph representation above: inputs, of type \texttt{xla::device\_data}, are associated with a device (which in this case is \texttt{CPU:0}).
All operations require all inputs are on the same device address space and subsequently their outputs stay on the same device.
While at first glance this choice appears to place a big burden on the user, frameworks offer ways to transfer entire models to a device with one or two lines of code.
In addition, this choice prevents difficult to debug performance problems associated with implicit transfers between device address spaces which the user didn't request.

\subsection{Control Flow}

\begin{figure}
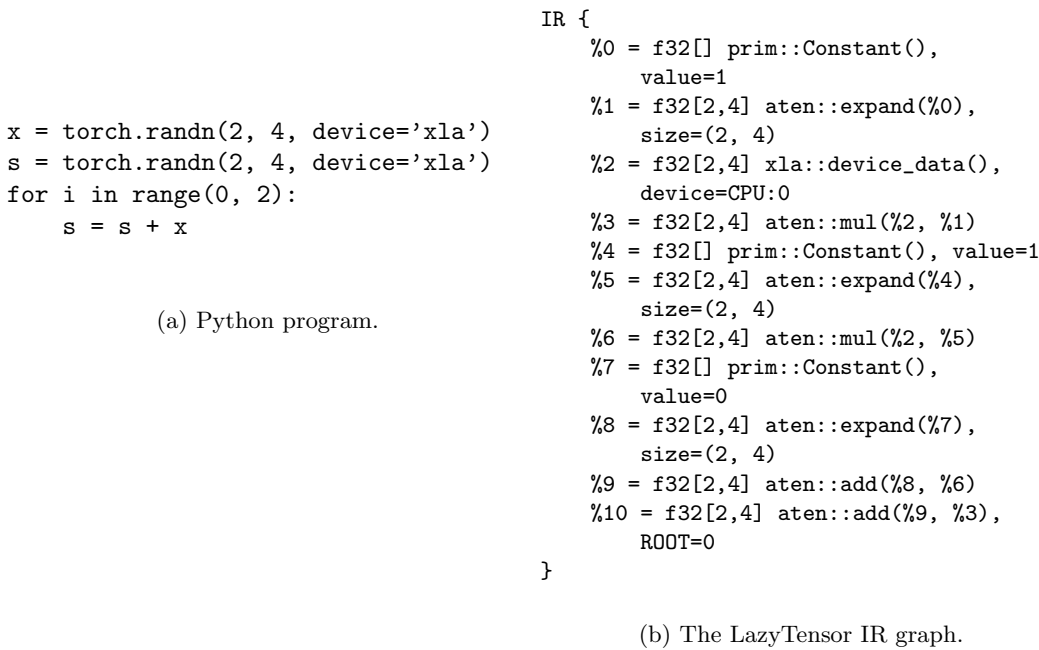

\centering
\begin{subfigure}[b]{0.45\textwidth}
    \begin{verbatim}
x = torch.randn(2, 4, device='xla')
s = torch.randn(2, 4, device='xla')
for i in range(0, 2):
    s = s + x
    \end{verbatim}
    \caption{Python program.}
    \label{fig:loopprog}
\end{subfigure}
\begin{subfigure}{0.45\textwidth}
    \begin{small}
    \begin{verbatim}
IR {
    %0 = f32[] prim::Constant(),
        value=1
    %1 = f32[2,4] aten::expand(%0),
        size=(2, 4)
    %2 = f32[2,4] xla::device_data(),
        device=CPU:0
    %3 = f32[2,4] aten::mul(%2, %1)
    %4 = f32[] prim::Constant(), value=1
    %5 = f32[2,4] aten::expand(%4),
        size=(2, 4)
    %6 = f32[2,4] aten::mul(%2, %5)
    %7 = f32[] prim::Constant(),
        value=0
    %8 = f32[2,4] aten::expand(%7),
        size=(2, 4)
    %9 = f32[2,4] aten::add(%8, %6)
    %10 = f32[2,4] aten::add(%9, %3),
        ROOT=0
}\end{verbatim}
    \end{small}
    \caption{The LazyTensor IR graph.}
    \label{fig:loopunroll}
\end{subfigure}
\caption{A simple program with a loop.}
\label{fig:loop}
\end{figure}

Currently, there is no representation for control flow in LazyTensor IR graphs.
For both conditionals and loops, the resulting execution path is captured and executed.
Given a simple program with a loop (Figure~\ref{fig:loopprog}), an unrolled, linear IR graph is generated (see Figure~\ref{fig:loopunroll}).

This reflects the separation between ``tensor programs'' and the host language (Python or Swift in our case).
While extending the system with control flow would be possible, generating IR graphs with control flow would require a degree of integration with the host language.
Moreover, our implementation works well in practice.
For example, conditional statements that select between a small number of model configurations or a loop that constructs a model out of repeated layers are both well served by this choice. On the other hand, dynamic conditional statements that branch on runtime \tensor\ values simply cause a break in the trace, rather than making tracing impossible.

\subsection{In-place operations}

Both \stf\ and PyTorch allow syntactically in-place updates: \texttt{s += x} updates the value of \texttt{s} to \texttt{s + x}.
Such operations are the cornerstone of weight updates during training and must be supported efficiently.

The two frameworks diverge in the way they support such operations.
Swift's operator overloads automatically convert the in-place version to the simple assignment \texttt{s = s + x} and therefore can reuse the regular addition implementation.
The XLA compiler will leverage the knowledge that the old value of \texttt{s} is no longer needed and can reuse memory efficiently.

On the other hand, Python doesn't offer such a rewrite and therefore PyTorch requires the implementation of additional, in-place versions of arithmetic (among other) operators.
However, the LazyTensor IR graphs have no concept of mutation.
Instead, all IR graphs represent pure functions.
This was chosen to accurately reflect the underlying XLA HLO IR.

To achieve mutation semantics, the LazyTensor system implements mutation as a substitution of the underlying computation associated with the destination with the computation associated with the expression on the right hand side of an in-place operation.
This achieves the same net effect as the built-in mechanism in Swift: the generated IR graph looks the same for regular and in-place operations.
We rely on runtime indirection to replace the underlying computation to achieve the desired semantics.

While mutation semantics can be achieved in a purely functional manner, memory usage and performance are crucial when training machine learning models.
A naive implementation of this model would use twice the optimal amount of memory for updates, due to the need to store the right hand side before replacing the destination with it.
Fortunately, the purely functional representation we've chosen matches the model of XLA, which allows specifying aliasing between input and output buffers\footnote{Details at \url{https://www.tensorflow.org/xla/aliasing}.}.
Destinations of in-place operations that end up as parameters of a LazyTensor IR graph can be ``donated'' for reuse as outputs through this mechanism, solving the problem of memory usage.

However, use of aliasing must be limited to IR graphs terminated at the step barrier.
Consider the following pseudo-code sequence:
\begin{verbatim}
  # Assume Tensor a is materialized data
  a += 1
\end{verbatim}
If we use aliasing for the IR graph rooted at \texttt{a}, the following sequence would behave incorrectly:
\begin{verbatim}
  print(a)
  print(a)
\end{verbatim}
With aliasing enabled, both \texttt{print(a)} statements would increment the buffer \texttt{a}, unless we force the evaluation of updated value of \texttt{a} instead of growing the IR graph.

Additionally, we must make sure that all live tensors are part of the computation when we use aliasing.
Consider the following sequence:

\begin{verbatim}
  # tensor a is immediate data
  b = a + 2
  a += 1
  print(a)
  print(b)
\end{verbatim}
If \texttt{b} is not part of the same IR graph as \texttt{a}, it'll use the updated value of \texttt{a}, which is incorrect.

Because of these two caveats, we limit our use of aliasing to computations which meet the following two conditions:
\begin{itemize}
    \item The computation behind each output is evaluated and not allowed to grow any further.
    \item All live \tensor s are part of the computation.
\end{itemize}
Both conditions are met at the end of the training step when \texttt{mark\_step()} is called inside \texttt{optimizer.step()} in PyTorch, as we accumulate both the complete forward and backwards pass.
In practice, this technique is only necessary to minimize memory usage for gradient updates, which also happen at the end of the training step.

\subsection{In-place operations on views}

PyTorch offers views as an explicit mechanism to control the sharing of underlying storage between \tensor s.
Several operations in its API return results which are guaranteed to share their underlying storage with their base \tensor s.
For example, in the following snippet of code:

\begin{verbatim}
  t = torch.rand(4, 4)
  v = t.view(2, 8)
\end{verbatim}

\texttt{v} is a reshape to size \texttt{(2, 8)} guaranteed to share its storage with \texttt{t}.
There are several other operations besides \texttt{view} itself which guarantee sharing semantics, some of which only operate on a subset of the storage (e.g. \texttt{narrow}, \texttt{permute}, etc).

In a purely functional paradigm, this feature has no implications as long as user programs respects referential transparency.
However, PyTorch allows in-place updates on views, which are guaranteed to be reflected in the base \tensor\ as well.
For example, \texttt{v.add\_(1)} will also update \texttt{t} since \texttt{v} is a view of \texttt{t}.

Our system supports this feature correctly, extending the approach used for in-place operations on regular tensors.
An update of a view creates two sequences of operations (still in the same IR graph):

\begin{enumerate}
    \item The ``forward'' sequence, which creates the computation required to represent the updated view. Multiple view operations can be applied to the base \tensor\ and we need to iterate through all.
    \item The ``backward'' sequence, which creates the computation required to represent the updated base. We start from the updated value of the view, iterate the view operations in reverse order and apply the inverse of each view operation.
\end{enumerate}

For example, if \texttt{x} is a tensor of shape \texttt{(2, 3, 4)}, \texttt{v = x.permute(1, 2, 0).add\_(42)} leads to the IR graph in Figure~\ref{fig:inplace}.

\begin{figure}
\begin{verbatim}
  IR {
      %0 = s64[] prim::Constant(), value=1
      %1 = s64[] xla::device_data(), device=CPU:0
      %2 = s64[] aten::mul(%1, %0)
      %3 = f32[2,3,4] xla::device_data(), device=CPU:0
      %4 = f32[3,4,2] aten::permute(%3), dims=(1, 2, 0)
      %5 = f32[3,4,2] aten::add(%4, %2)
      %6 = f32[2,3,4] aten::permute(%5), dims=(2, 0, 1), ROOT=0
  }
\end{verbatim}
\caption{A IR graph involving in-place operations.}
\label{fig:inplace}
\end{figure}

The resulting IR graph contains both the permutation directly specified by the user---in this case \texttt{(1, 2, 0)} and its inverse, \texttt{(2, 0, 1)}.
The former is used to compute the underlying computation which represents \texttt{v}, while the latter is used for representing the updated value of \texttt{x}, the source of the view.

\subsection{Unimplemented \tensor\ Operations}

Machine learning frameworks offer thousands of operations in their \tensor\ APIs.
Although DSCs often support more flexible linear algebra operations than libraries of pre-compiled kernels, DSCs often do not support all \tensor\ operations.
Examples include image decompression algorithms.
In order to deliver a \emph{drop-in} replacement \tensor\ API, all \tensor\ operations that do not have lowerings to the DSC are implemented with the following pattern:
\begin{enumerate}
    \item Evaluate the IR graph for all operation inputs.
    \item Call the underlying eager implementation on the evaluated inputs.
    \item Start a new IR graph with the eager op execution's result.
\end{enumerate}
Thus, every \tensor\ operation available on the type will result in a correct program when executed with the LazyTensor system, albeit with some potential performance implications. %
Using a debugger's breakpoint facility, users can quickly determine where their code calls operations with no implemented DSC lowering.

\section{Evaluation}
\label{section:evaluation}

We evaluate the LazyTensor system across a number of dimensions and applications.

\subsection{Code Reuse}

The core of the LazyTensor system has been used to power XLA integrations across both PyTorch and Swift for TensorFlow.
Table~\ref{tab:sloc} documents the source lines of code (SLoC) within the respective folders of the implementations in Swift for Tensorflow, and annotates whether they are shared between the two frontends.
Because this measure is somewhat crude, we estimate slightly above 75\% of the SLoC are shared, despite the folder-based method of estimation indicating around 85\% SLoC reuse.
This demonstrates this technique---and a substantial fraction of the implementation---is reusable across programming languages, and underlying eager runtimes.

\begin{table}[h]
    \centering
    \begin{tabular}{c|r|c}
        Component Folder & SLoC & Shared? \\
        \hline
        \texttt{xla\_tensor} \& \texttt{xla\_client} & 49,910 & Yes \\
        \texttt{swift\_bindings} & 7,325 & Swift-only \\
        \texttt{CX10} & 1,697 & Swift-only \\
    \end{tabular}
    \caption{Source lines of code (SLoC) in Swift for Tensorflow used to implement individual components of the system as grouped by folder, and whether they are shared between the Swift and Python implementations of the LazyTensor technique.}
    \label{tab:sloc}
\end{table}

\subsection{Training Transformers on Cloud TPUs}

\begin{table*}
    \centering
    \begin{tabular}{|l|c|c|c|r|}
        \hline
        Framework & Accelerator & Global Batch Size & Eval Perplexity & Training Time \\
        \hline
        PyTorch Native & 4x V100 & 48 & 3.14 & 133.4 minutes \\
        \hline
        \multirow{2}{*}{PyTorch LazyTensor} & Cloud TPU v3-8 & \multirow{2}{*}{128} & \multirow{2}{*}{3.25} & \multirow{2}{*}{38.3 minutes} \\
        & {\scriptsize (4x TPUv3 chips)} & & & \\
        \hline
    \end{tabular}
    \caption{Time required to train HuggingFace's \texttt{roberta-base} (125M params) on the raw WikiText-103 dataset~\cite{wikitext} for 3 epochs using half precision. The largest batch size that was able to fit in each chip's memory was used for above numbers to maximize utilization and thus optimize training time.}
    \label{tab:huggingfaceperf}
\end{table*}

The Transformer~\cite{attention} is a deep learning architecture widely used in the natural language processing domain today, which has resulted in state of the art performance on various metrics everywhere from language parsing, machine translation, to question answering.
This architecture has heralded a wide lineage of attention based models such as BERT~\cite{devlin-etal-2019-bert}, T5~\cite{2019t5}, and more.

With the PyTorch LazyTensor implementation, we've enabled the popular HuggingFace Transformer library~\cite{huggingfaces} to run on Cloud TPUs using XLA.
PyTorch with LazyTensor was able to demonstrate significant performance improvements on TPUs compared to roughly equivalent GPU hardware (Table~\ref{tab:huggingfaceperf}).

\subsection{Scaling ResNet-50 on TPUs}

We evaluate the scaling properties of the LazyTensor system by using \stf\ to train ResNet-50~\cite{he2016identity} on TPU Pods.
The performance of \stf\ on TPUs was measured by training the ResNet-50 image classification network~\cite{he2016identity} on the ImageNet 2012 dataset~\cite{imagenet_cvpr09} using TPUv3-16, TPUv3-32 and TPUv3-128 clusters (not using Cloud), shown in Table~\ref{tab:s4tftpuresnetperf}.
The model was trained for 90 epochs, and both the time required as well as the post-warmup throughput in examples per second were recorded.
Per-accelerator throughput is substantially maintained, demonstrating that the LazyTensor technique can scale to large TPU super-computers.

\begin{table*}[h]
    \centering
    \begin{tabular}{|l|r|r|r|r|}
        \hline
        \# Cores & Validation Accuracy & Training Time & Throughput & Per-Accelerator Throughput \\
        & (top-1) & (90 epochs) & (examples / s) & (examples / s / TPU core) \\
        \hline
        16 & 78.1\% & 189 minutes & 10164 & 635.25 \\
        32 & 77.7\% & 96 minutes & 20015  & 625.47 \\
        128  & 77.8\% & 25 minutes & 77726  & 607.23 \\
        \hline
    \end{tabular}
    \caption{
        \stf\ training performance for ResNet-50 on ImageNet on TPUv3 clusters.
        Per-accelerator throughput is largely maintained while scaling from a single host to 8 hosts synchronously training a single model in data-parallel fashion, demonstrating the scalability of the LazyTensor approach employed by \stf\ to target TPUs.
    }
    \label{tab:s4tftpuresnetperf}
\end{table*}

\begin{table*}
    \centering
    \begin{tabular}{|c|r|r|c|}
        \hline
        Operation \& Length & Eager (ms) & LazyTensor (ms) & Ratio \\
        \hline
        Score 4 & 96 & 65 & 0.68 \\
        Score 8 & 182 & 118 & 0.65 \\
        Score 14 & 282 & 165 & 0.59 \\
        \hline
        Score \& Gradient 4 & 425 & 281 & 0.66 \\
        Score \& Gradient 8 & 894 & 581 & 0.65\\
        Score \& Gradient 14 & 1350 & 853 & 0.63 \\
        \hline
    \end{tabular}
    \caption{Time spent in respective operations of the WordSeg algorithm \cite{wordseg} based on the different \stf\ \tensor\ approaches when run on a GPU.}
    \label{tab:wordseg}
\end{table*}

\subsection{Limitations of XLA compilation}

Unfortunately, not all models achieve higher performance using the LazyTensor system as compared to the eager-system equivalent.
If programs do not run for long enough, the advantages of specialization do not outweigh the overheads of the JIT-compilation itself.
As a result, this approach often only makes sense during long-running, iterated computations such as neural network training, or batch inference.

One of the most painful limitations comes not from the technique itself, but the underlying DSC: XLA's \emph{static shape} limitation.
All \tensor\ shapes must be known at IR graph compile time, as they are used for static memory planning and other optimizations within the XLA compiler.
Although the system caches XLA programs based on a canonicalization of the LazyTensor IR graph, some ML models never converge to a ``shape stable'' set of IR graphs.
For example, training MaskRCNN~\cite{maskrcnn} on the COCO~\cite{DBLP:journals/corr/LinMBHPRDZ14} dataset displays poor accelerator utilization, as the accelerator remains idle while XLA repeatedly compiles new shape-specialized accelerator programs.
This application validated our choice to faithfully implement an \emph{identical API} to the original eager execution, as it enables users to pick whichever execution strategy works most effectively for their applications at any given time.

\section{Related work}
\label{section:related-work}

JAX~\cite{jax, jaxtracing} also builds on top of XLA, but uses explicit tracing--versus LazyTensor's implicit traces--to build optimizable program representations in their jaxpr IR.
This tracing approach forces users to refactor their code if any single function in the user's program calls some un-traceable functionality such as a black-box environment or uses non-\tensor-based data structures in a data-dependent way.
Although DyNet~\cite{dynet} builds traces, and performs optimizations including automatic batching on the trace IR, it still dispatches ops in the trace eagerly via cuDNN~\cite{cudnn} on GPUs and Eigen/MKL on CPUs instead of using a domain specific JIT compiler.

TensorFlow 1.X~\cite{tensorflow} users build an explicit graph data structure (\texttt{GraphDef}) using Python-based metaprogramming.
The extra indirection inhibits debugging, and makes combining \tensor--based and non-\tensor--based algorithms more difficult.
Although TensorFlow does have support for ``partial run'' to allow resuming the computational graph, this is feature is incompatible with TensorFlow's XLA support.
TensorFlow 2 \cite{tensorflow-eager} avoids the explicit meta-programming of graph-building, but like JAX requires entire functions and all functions they transitively call to be traced and lifted into the IR.
Taichi~\cite{taichi} and Numba~\cite{numba} similarly JIT-compile subsets of Python.
Our approach works across multiple languages, and allows for ergonomic mixing of code that cannot be optimized with the domain specific compiler.

Julia's XLA support~\cite{juliaTPU} translates Julia IR to XLA HLO without the need for tracing.
Unfortunately, because XLA does not support dynamic memory allocation, the subset of the Julia program translated to XLA must equivalently not use dynamic memory allocation.
It is this restriction that caused \stf\ to abandon this architecture (similar techniques are there called graph program extraction) and instead adopt the LazyTensor approach.
Fortunately, the Julia language has a number of other meta-programming mechanisms that combine with Julia's JIT-based specializing runtime to generate families of programs, effectively working around limitations induced by XLA's static memory model.

\section{Future work}
\label{section:future-work}

Although the LazyTensor system has been effectively employed in multiple applications, there are a number of directions of improvement to broaden its applicability.

Despite LazyTensor graph construction overheads not affecting time-to-convergence for most training workloads, updating the IR graph can affect workloads operating on small models, small tensor sizes or low batch sizes, as is often the case for inference. While our system skips expensive recompilation when encountering the same IR graph for a second time, such workloads could benefit from mitigation of graph construction overheads as well.

One possible direction would address the representation of the IR graphs. Reduction of IR node size and usage of data structures and algorithms which minimize pointer chasing have been successful strategies in more traditional just-in-time compilers such as WebKit's FTL JIT \footnote{https://webkit.org/blog/5852/introducing-the-b3-jit-compiler/}. We could borrow such techniques in our system to expand its appeal to workloads which would require a lower overhead in maintaining the IR graphs.

Another direction would provide users with a way to guarantee that the underlying computation of a \tensor\ is going to remain the same across steps. In doing so, our system would be allowed to skip the IR graph construction as well, after constructing it in the usual way during the first iteration.

Further, allowing users to encode control flow in the optimized program through special program annotations could increase performance in certain circumstances such as data-dependent branching.

Finally, techniques to automatically truncate, re-roll loops and asynchronously dispatch IR graph fragments could eliminate the need for \ltb\ in user code and mitigate re-compilations due to variable upper bounds for loops. However, doing so efficiently could require some degree of cooperation with the host language implementation. For example, while recognizing an unrolled loop pattern is possible entirely in the LazyTensor system, the host language could instead provide hints about the presence of such a pattern.

\section{Conclusion}
\label{section:conclusion}

In this paper we've introduced LazyTensor, a general technique to combine eager execution with domain specific compilers that does not restrict the expressivity of the user's programming language.

We have successfully implemented this technique in two programming languages for two machine learning frameworks: PyTorch and Swift for TensorFlow. We managed to reuse the majority of the implementation while targeting completely different languages and \tensor\ APIs. 

Our evaluation shows that we can efficiently target hardware that is only accessible via a DSC (XLA), which can result in significant performance improvements as measured by the HuggingFace Transformers library on Cloud TPUs.
Additionally, we have shown how this approach can scale to large distributed accelerator clusters.

\bibliographystyle{plain}
\bibliography{references}
\end{document}